\documentclass[prd,twocolumn,showpacs,showkeys,preprintnumbers,amsmath,amssymb,floatfix]{revtex4}

\usepackage[utf8]{inputenc}
\usepackage{hyperref}

\usepackage{dcolumn}
\usepackage{bm}
\usepackage{graphicx}
\usepackage{subfigure}

\def\be{\begin{equation}}
\def\ee{\end{equation}}
\def\bestar{\begin{equation*}}
\def\eestar{\end{equation*}}

\begin{document}

\title{Noncommutative Dirac quantization condition using the Seiberg-Witten map}

\author{Marco Maceda}
\email{mmac@xanum.uam.mx}
\affiliation{Departamento de F\'{i}sica \\ Universidad Aut\'{o}noma Metropolitana - Iztapalapa\\
                   Av. San Rafael Atlixco 186, A.P. 55-534, C.P. 09340, M\'exico D.F., M\'exico}
                   
\author{Daniel Martínez-Carbajal}
\email{leinadmar10@yahoo.com}
\affiliation{Departamento de F\'{i}sica \\ Universidad Aut\'{o}noma Metropolitana - Iztapalapa\\
                   Av. San Rafael Atlixco 186, A.P. 55-534, C.P. 09340, M\'exico D.F., M\'exico}

\date{\today}

\begin{abstract}

We investigate the validity of the Dirac quantization condition (DQC) for magnetic monopoles in noncommutative space-time. We use an approach based on an extension of the method introduced by Wu and Yang; the effects of noncommutativity are analyzed using the Seiberg-Witten map and the corresponding deformed Maxwell's equations are discussed. By using a perturbation expansion in the noncommutativity parameter $\theta$, we show first that the DQC remains unmodified up to the first and second order. This result is then generalized to all orders in the expansion parameter for a class of noncommutative electric currents induced by the Seiberg-Witten map; these currents reduce to the Dirac delta function in the commutative limit.

\end{abstract}


\maketitle

\section{Introduction}
\label{intro}

Magnetic monopoles were suggested originally as a source for symmetry between the electric and magnetic fields being at the same time compatible with quantum mechanics;  they lead to the quantization condition  of the electric charge in terms of the charge of a magnetic monopole now known as Dirac's quantization condition~\cite{Dirac:1931kp}. Several features of magnetic monopoles have been generalized to the context of non-Abelian gauge theories, where the 't Hooft-Polyakov monopole~\cite{'tHooft:1974qc,Polyakov:1974ek} gives a configuration that for long distances from the source reduces to Dirac's solution.

The analysis of magnetic monopoles has contributed to the development not only of mathematical tools, but also to the understanding of at first sight unrelated systems. More recently, monopole-like structures have appeared in different contexts such as superfluids and Bose-Einstein condensation; experimental work has permitted to built systems with properties analogue to them. 

In the formulation due to Dirac, a nodal singularity (Dirac's string) is present since the gauge potential of the electromagnetic field is ill-defined along it; this can be reformulated as the condition that the wave function of a particle moving in this field should have a non-integrable phase~\cite{Dirac:1931kp,Goddard:1977da}.  

To avoid the presence of singularities, Wu and Yang~\cite{PhysRevD.12.3845} introduced two coordinates charts for the magnetic monopole and separated its gauge potential accordingly; the resulting gauge potentials are then regular in their respective domains and can be connected by a non-singular gauge transformation in an overlapping region.  

In~\cite{salminen2011magnetic,LangvikSalminenTureanu2011} the extension of this construction to the noncommutative framework was considered. For this purpose noncommutative Maxwell's equations with group $U_*(1)$ were written using a star product and the associated gauge potentials and electromagnetic fields were derived in a perturbative treatment in terms of powers of the noncommutative parameter. Dirac's quantization condition was shown to fail to second order in perturbation theory.

In this work we investigate the validity of Dirac's quantization condition using the Seiberg-Witten (SW) map~\cite{Seiberg:1999vs}. As it is well-known, this map allows a straightforward construction of a noncommutative gauge theory from a commutative one, the basic ingredients being the knowledge of the commutative gauge potentials, gauge parameter and matter fields. Over the years a lot of effort has been put into obtaining closed expressions for the noncommutative fields at arbitrary order in the noncommutative parameter; fortunately enough, an iterative procedure that completely solves the SW map is known~\cite{Ulker:2007fm,Ulker:2012yk}.

Starting from the gauge potential of Wu and Yang, the SW map will allow us to obtain explicit expressions for the noncommutative gauge potentials to arbitrary order on the noncommutative parameter. A general Ansatz for the gauge potentials of the noncommutative monopole field can be guessed; we exploit their symmetries to investigate the noncommutative corrections to the classical gauge parameter. It is important to mention that the potentials obtained by the SW map are non-singular in their domains of definition. It turns out that the noncommutative corrections can be calculated explicitly by an iterative procedure and they can be shown to vanish. Dirac's quantization condition is then preserved in this scheme.

The knowledge of these potentials can be used in turn to write down modified Maxwell's equations and from them, by following a similar procedure as in~\cite{salminen2011magnetic,LangvikSalminenTureanu2011}, we deduce the noncommutative Ampère and Gauss laws. Once this is achieved, we use these equations to deduce possible sources for the electromagnetic field that may arise due to noncommutativity instead of imposing a structure on them from the very beginning. More specifically, we let open the possibility that an electric current may be present into the noncommutative Maxwell equations. 

The paper is organized as follows: In Sec.~\ref{secc:2} we review the construction of noncommutative gauge theories using the SW map. The general expressions for the noncommutative gauge fields to arbitrary order on the noncommutative parameter are then discussed in Sec.~\ref{secc:3}. The classical DQC is then discussed in Sec.~\ref{secc:4} and in Sec.~\ref{secc:5} noncommutative gauge transformations are analyzed. Modified Maxwell's equations are formulated in Sec.~\ref{secc:6} and the induced sources are identified. The solution to Maxwell's modified equations involving the noncommutative gauge potential and gauge parameter is then given in Sec.~\ref{secc:7}; there it is shown that the noncommutative corrections to the gauge parameter vanish and hence the DQC is preserved. We finally end with our conclusions. 

We shall use units in which $\hbar=c=g=e=1$ throughout, unless otherwise stated.

\section{The Seiberg-Witten Map }
\label{secc:2}

\subsection{Noncommutative Gauge Theories }

Noncommutative gauge theories need noncommutative gauge fields to define covariant derivatives. If we have a matter field transforming as $\delta_{\lambda}\Phi=i\lambda*\Phi$, then 
\begin{equation}
D_{\mu}\Phi=\partial_{\mu}\Phi-iA_{\mu}*\Phi,
\label{eq:derivative covariant}
\end{equation}
defines a covariant derivative if the gauge field $A_{\mu}$ transforms according to the rule
\begin{equation}
\delta_{\lambda}A_{\mu}=\partial_{\mu}\lambda+i[\lambda,\,A_{\mu}]_{*}.
\end{equation}
Here $[A,\,B]$ denotes the Moyal-Groenewold bracket $[A,\,B] :=A*B-B*A$. Similarly, we can define
\begin{equation}
F_{\mu\nu} := \partial_{\mu}A_{\nu}-\partial_{\nu}A_{\mu}-i[A_{\mu},\,A_{\nu}]_{*},
\label{eq:the field strength}
\end{equation}
as the field strength with transformation law $\delta_{\lambda}F_{\mu\nu}=i[\lambda,\,F_{\mu\nu}]_{*}$. In this case the covariant derivative Eq.~(\ref{eq:derivative covariant}) is compatible with the gauge transformation 
\begin{equation}
A_{\mu}\rightarrow A_{\mu}^{\prime}=U*A_{\mu}*U^{-1}+iU*\partial_{\mu}U^{-1},
\end{equation}
where $U := e^{i\lambda (x)}_*$. Therefore, from any commutative gauge theory as a starting point, we could construct a noncommutative  one by substitution of the usual product of functions by the Moyal product. The noncommutative invariant action for the gauge sector is then
\begin{eqnarray}
S=\int d^{4}x \, F^{\mu\nu}*F_{\mu\nu}.
\end{eqnarray}
However, for a Lie group $G$, with corresponding Lie algebra $\mathcal{G}$ generated by $n$ elements $\{T_{a}\}$ satisfying $[T_{a},\,T_{b}]=f_{ab}^{\:\:c}T_{c}$, we have in general
\begin{eqnarray}
[\lambda,\,A_{\mu}]_{*} & = & \left(\lambda^{a}*F_{\mu\nu}^{b}-F_{\mu\nu}^{a}*\lambda^{b}\right)T_{a}T_{b}
\nonumber \\[4pt]
 & = & \frac{1}{2}\left(\lambda^{a}*F_{\mu\nu}^{b}-F_{\mu\nu}^{a}*\lambda^{b}\right)[T_{a},\,T_{b}]
 \nonumber \\[4pt]
 & + & \frac{1}{2}\left(\lambda^{a}*F_{\mu\nu}^{b}-F_{\mu\nu}^{a}*\lambda^{b}\right)\{T_{a},\,T_{b}\},
\end{eqnarray}
where $A_{\mu}=A_{\mu}^{a}T_{a}$ and $\lambda=\lambda^{a}T_{a}$ with $a=1, \dots, n$. These gauge transformations generate components in the enveloping algebra $\mathcal{U}$ of $\mathcal{G}$ obtained from all the products of $\mathcal{G}$. Since
\begin{equation}
T_{a}T_{b}=\frac{1}{2}[T_{a},\,T_{b}]+\frac{1}{2}\{T_{a},\,T_{b}\},
\end{equation}
the enveloping algebra can be obtained by repeatedly computing all commutators and anticommutators until it closes assuming that in general we can write
\begin{equation}
[T_a,\,T_b]=if_{ab}^{c}T_{c},\qquad \{T_{a},\,T_{b}\}=d_{ab}^{c}T_{c}.
\end{equation}
An example of the above relations is given by the Lie algebras of the group $U(n)$ where in the fundamental representation they coincide with their enveloping algebras. A Lie algebra coincides with its enveloping algebra since this depends on the representation. For instance, in the case of $SU(2)$ in the fundamental representation, the generators are the Pauli matrices, satisfying $[\sigma_{a},\,\sigma_{b}]=i\varepsilon_{abc}\sigma_{c},\;\{\sigma_{a},\,\sigma_{b}\}=2\delta_{ab}I$. Thus, the enveloping algebra contains the unit matrix besides the Pauli matrices, i.e. it corresponds to $U(2)$. For the vector representation, the generators are $(T_{a})_{b}^{\:c}=i\varepsilon_{ab}^{\;\:c}$, and it can be shown that its enveloping algebra is given then by $U(3)$. This means that the number of degrees of freedom of a noncommutative theory is higher than that of a commutative one. Nevertheless, the number of gauge parameters will also increase, implying that some of the new degrees of freedom can be gauged away; the Seiberg-Witten map is such that the number of degrees of freedom is the same in both commutative and noncommutative gauge theories.

\subsection{The Seiberg-Witten map }

String theory points out to a relation between standard gauge theories and noncommutative ones in terms of a gauge equivalence relation dictated by the Seiberg-Witten map~\cite{Seiberg:1999vs}
\begin{equation}
\widehat{A}_{\mu}\left(A+\delta_{\lambda}A;\,\theta\right)=\widehat{A}_{\mu}\left(A;\,\theta\right)+\widehat{\delta}_{\widehat{\Lambda}}\widehat{A}_{\mu}\left(A;\,\theta\right),\label{eq:equivalence relation}
\end{equation}
where $A$ and $\lambda$ are the standard gauge field and gauge parameter respectively; this is the analog of the ordinary gauge transformation
\begin{equation}
\delta_{\lambda}A_{\mu}=\partial_{\mu}\lambda+i[\lambda,\,A_{\mu}]=D_{\mu}\lambda.\label{eq: noncommutative gauge transformations A}
\end{equation}
We can rewrite Eq.~(\ref{eq:equivalence relation}) as
\begin{equation}
\widehat{\delta}_{\widehat{\varLambda}}\widehat{A}_{\mu}\left(A;\,\theta\right)=\widehat{A}_{\mu}\left(A+\delta_{\lambda}A;\,\theta\right)-\widehat{A}_{\mu}\left(A;\,\theta\right)=\delta_{\varLambda}\widehat{A}_{\mu}\left(A;\,\theta\right).
\label{eq:Seiberg-Witten map-A transformation}
\end{equation}
The ordinary gauge transformation on the r.h.s. of Eq.~(\ref{eq:Seiberg-Witten map-A transformation}) acts on the components of $\widehat{A}$ when it is expanded as a power series in $\theta$ and the NC gauge field $\widehat{A}$ and NC gauge parameter $\widehat{\varLambda}$ are assumed to have the following functional dependence \cite{Seiberg:1999vs}
\begin{equation}
\widehat{A}_{\mu}=\widehat{A}_{\mu}\left(A;\,\theta\right),\quad\widehat{F}_{\mu\nu}=\widehat{F}_{\mu\nu}\left(A;\,\theta\right),\quad\widehat{\varLambda}=\widehat{\varLambda}_{\lambda}\left(\lambda,\,A;\,\theta\right).
\end{equation}
It should be noted that Eq.~(\ref{eq:Seiberg-Witten map-A transformation}) can be implemented for linear and adjoint representations.

The Seiberg-Witten map is a tool to construct noncommutative gauge theories having an explicit dependence on the commutative fields and their derivatives; it has the characteristic feature that the number of degrees of freedom of the original theory is preserved. For a noncommutative gauge theory constructed in this way we have
\begin{equation}
\delta_{\lambda}S=\widehat{\delta}_{\widehat{\varLambda}}S=0.
\end{equation}
This result implies invariance of the action; it can be understood either in terms of the noncommutative fields, with associated noncommutative gauge transformations Eq.~(\ref{eq: noncommutative gauge transformations A}), or directly in terms of transformations involving the commutative fields. It is known that the map can be written for any gauge group~\cite{wess2001non,Jurco:2000ja,Madore:2000en,Jurco:2001my,Jurco:2001rq}, and it can be solved iteratively~\cite{Ulker:2007fm}. The first step on the solution consists in writing the fields as a power series on the noncommutative parameters,
\begin{equation}
\widehat{A}_{\mu}=A_{\mu}^{0}+A_{\mu}^{1}+A_{\mu}^{2}+\cdots.
\label{eq:noncommutative A}
\end{equation}
 
Eq.~(\ref{eq:equivalence relation}) should then be solved simultaneously for $\widehat{A}_{\mu}$ and $\widehat{\varLambda}_{\lambda}$ and this can be cumbersome especially when looking for higher order solutions in $\theta$. In general, the commutative parameters $\lambda$ associated to a linear representation, where $\delta_{\lambda}\Phi=i\lambda\Phi=i\lambda^{a}T_{a}\Phi$, satisfy the following classical cocycle condition
\begin{eqnarray}
[\delta_{\alpha},\,\delta_{\beta}]\Phi=-[\alpha,\,\beta]\Phi=\delta_{-i[\alpha,\,\beta]}\Phi.
\end{eqnarray}
or equivalently, the ordinary gauge consistency condition
\begin{equation}
\delta_{\alpha}\delta_{\beta}-\delta_{\beta}\delta_{\mathbf{\alpha}}=\delta_{-i[\alpha,\,\beta]}.
\end{equation}
The noncommutative parameters must depend on the commutative gauge fields, $\widehat{\varLambda}=\widehat{\varLambda}_{\lambda}\left(\lambda,\,A;\,\theta\right)$ and in analogy with the commutative case we can write
\begin{equation}
\delta_{\alpha}\hat \Phi=i\Lambda_\alpha * \widehat{\Phi}.
\end{equation}
In consequence we have 
\begin{eqnarray}
\delta_{\alpha}\delta_{\beta}\widehat{\Phi} & = & \delta_{\alpha}\left(\widehat{\delta}_{\widehat{\beta}}\widehat{\Phi}\right)=i\delta_{\alpha}\widehat{\varLambda}_{\beta}*\widehat{\Phi}+i\widehat{\varLambda}_{\beta}*\delta_{\alpha}\widehat{\Phi},
\nonumber \\[4pt]
 & = & i\delta_{\alpha}\widehat{\varLambda}_{\beta}*\widehat{\Phi}-\widehat{\varLambda}_{\beta}*\widehat{\varLambda}_{\alpha} *\widehat{\Phi},
\end{eqnarray}
or
\begin{eqnarray}
[\delta_{\alpha},\,\delta_{\beta}]\widehat{\Phi} & = & (i\delta_{\alpha}\widehat{\varLambda}_{\beta}-i\delta_{\beta}\widehat{\varLambda}_{\mathbf{\alpha}} + [\widehat{\varLambda}_{\mathbf{\alpha}},\,\widehat{\varLambda}_{\beta}]_{*}) *\widehat{\Phi}.
\end{eqnarray}
Hence, we have the transformation law for the noncommutative parameters,
\begin{equation}
i\delta_{\alpha}\widehat{\varLambda}_{\beta}-i\delta_{\beta}\widehat{\varLambda}_{\mathbf{\alpha}}+[\widehat{\varLambda}_{\mathbf{\alpha}},\,\widehat{\varLambda}_{\beta}]_{*} = i\widehat{\varLambda}_{\mathbf{-i[\alpha,\,\beta]}}.
\end{equation}
In order to solve this equation, we write a series development
\begin{equation}
\widehat{\varLambda}_{\lambda}=\lambda+\varLambda_{\lambda}^{1}+\varLambda_{\lambda}^{2}+\cdots.
\end{equation}
The solution to first order $\varLambda^{1}$ is~\cite{Seiberg:1999vs},
\begin{equation}
\hat{\varLambda}(\lambda,\:A)=\lambda+\frac{1}{4}\theta^{\mu\nu}\{\partial_{\mu}\lambda,\:A_{\nu}\}+\mathcal{O}(\theta^{2}).\label{eq:Lamda first orderS-W}
\end{equation}
Inserting this solution into Eq.~(\ref{eq:Seiberg-Witten map-A transformation}) gives
\begin{equation}
\hat{A}_{\xi}(A)=A_{\xi}-\frac{1}{4}\theta^{\mu\nu}\{A_{\mu},\:\partial_{\nu}A_{\xi}+F_{\nu\xi}\}+\mathcal{O}(\theta^{2}),
\label{eq:potential A first orderS-W}
\end{equation}
and the associated field strength has the form
\begin{eqnarray}
\widehat{F}_{\gamma\rho} &=& F_{\gamma\rho}-\frac{1}{4}\theta^{kl}\left(\{A_{k},\,\partial_{l}F_{\gamma\rho}+D_{l}F_{\gamma\rho}\}-2\{F_{\gamma k},\,F_{\rho l}\}\right)
\nonumber \\[4pt]
&&+\mathcal{O}(\theta^{2}).
\label{eq: field strength first order S-W}
\end{eqnarray}
The solution for matter fields to first order is then
\begin{equation}
\widehat{\Phi}=\Phi+\frac{1}{2}\theta^{\mu\nu}\left(-A_{\mu}\partial_{\nu}\Phi+\frac{1}{2}A_{\mu}A_{\nu}\Phi\right).
\end{equation}
in the fundamental representation. On the other hand, for the adjoint representation the equation to be solved
is $\delta_{\lambda}\Phi=i[\widehat{\varLambda},\,\widehat{\Phi}]_{*}$; the solution is
\begin{equation}
\widehat{\Phi}=\Phi-\frac{1}{4}\theta^{\mu\nu}\{A_{\mu},\,\left(D_{\nu}+\partial_{\nu}\right)\Phi\}+\mathcal{O}(\theta^{2}).
\end{equation}
Higher order terms can be obtained in the same way, or by use of the equation~\cite{Seiberg:1999vs}
\begin{equation}
\frac{\partial}{\partial\theta^{\mu\nu}}\widehat{\Phi}=\widehat{\Phi^{1}}_{\mu\nu},
\end{equation}
where $\widehat{\Phi^{1}}_{\mu\nu}$ is obtained from the first order term of the map by substituting the fields by their noncommutative
counterparts, all of them multiplied by the Moyal product.  

As mentioned before, the most general solution of the SW map has an infinite number of parameters. Depending on the problem at hand, some solutions may be better suited than others~\cite{wess2001non,Jurco:2000ja,Madore:2000en,Jurco:2001my,Jurco:2001rq}. A nice feature of the above solutions is that corrections to the field strength vanish if the commutative field strength vanishes.

\section{Seiberg-Witten maps to all orders \label{sec:Seiberg=002013Witten-Maps-toAll Orders}}
\label{secc:3}

In this section we review the main features of the solutions to the SW equations.

\subsection{First order solution}

In~\cite{Seiberg:1999vs} the first order solution was given as
\begin{eqnarray}
\varLambda_{\lambda}^{1} & = & \frac{1}{4}\theta^{\mu\nu}\{\partial_{\mu}\lambda,\:A_{\nu}\},\label{eq:parameter first order}
\\[4pt]
A_{\gamma}^{1} & = & -\frac{1}{4}\theta^{\mu\nu}\{A_{\mu},\:\partial_{\nu}A_{\gamma}+F_{\nu\gamma}\}.
\label{eq:SW Potential A first order}
\end{eqnarray}
The field strength is calculated as
\begin{equation}
F_{\gamma\rho}^{1}=-\frac{1}{4}\theta^{\mu\nu}\left(\{A_{\mu},\,\partial_{\nu}F_{\gamma\rho}+D_{\nu}F_{\gamma\rho}\}-2\{F_{\gamma\mu},\,F_{\rho\nu}\}\right).\label{eq:field strength-first order}
\end{equation}
We may rewrite this expression in terms of the first order potential $A_{\mu}^{1}$ and the commutative potential $A_{\mu}^{0}$. After some simplifications we obtain
\begin{equation}
\partial_{\gamma}A_{\rho}^{1}-\partial_{\rho}A_{\gamma}^{1}=-\theta^{\mu\nu}\left(A_{\mu}^{0}\partial_{\nu}F_{\gamma\rho}^{0}+\partial_{\mu}A_{\gamma}^{0}\partial_{\nu}A_{\rho}^{0}-F_{\gamma\mu}^{0}F_{\rho\nu}^{0}\right),
\end{equation}
where $F_{ik}^{0}=\partial_{i}A_{k}^{0}-\partial_{k}A_{i}^{0}$ is the field strength tensor at zero order. From this it follows that
\begin{equation}
F_{\gamma\rho}^{1}=\partial_{\gamma}A_{\rho}^{1}-\partial_{\rho}A_{\gamma}^{1}+\theta^{\mu\nu}\partial_{\mu}A_{\gamma}^{0}\partial_{\nu}A_{\rho}^{0}.
\label{fmunu1st}
\end{equation}
This rewriting will be useful in later calculations.

\subsection{Second order solution}

The second order solution of the SW map was given in~\cite{Moller:2004qq}; it can be recast  as~\cite{Ulker:2007fm}
\begin{eqnarray}
\varLambda_{\lambda}^{2} & = & -\frac{1}{8}\theta^{kl}\left(\{A_{k}^{1},\:\partial_{l}\lambda\}+\{A_{k},\:\partial_{l}\varLambda^{1}\}\right),\label{eq:parameter second order}
\\[4pt]
A_{\gamma}^{2} & = & -\frac{1}{8}\theta^{kl}\{A_{k}^{1},\:\partial_{l}A_{\gamma}^{0}+F_{l\gamma}^{0}\}+\{A_{k}^{0},\:\partial_{l}A_{\gamma}^{1}+F_{l\gamma}^{1}\}.
\label{eq:potential A second order}
\end{eqnarray}
In terms of first order solutions, the field strength at the second order~\cite{Moller:2004qq} can also be written as
\begin{eqnarray}
F_{\gamma\rho}^{2} &=& -\frac{1}{8}\theta^{\mu\nu}\left(\{A_{\mu}^{0},\,\partial_{\nu}F_{\gamma\rho}^{1}+(D_{\nu}F_{\gamma\rho}^{0})^{1}\} \right.
\nonumber \\[4pt]
&&+ \{A_{\mu}^{1},\,\partial_{\nu}F_{\gamma\rho}^{0}+D_{\nu}F_{\gamma\rho}^{0}\}-2\{F_{\gamma\mu}^{0},\,F_{\nu\rho}^{1}\}
\nonumber \\[4pt]
&&\left. -2\{F_{\gamma\mu}^{1},\,F_{\rho\nu}^{0} \} \right).
\label{eq:field strength-second order}
\end{eqnarray}
Here the covariant derivative $(D_{\nu}F_{\gamma\rho}^{0})^{1}$ is given by
\begin{equation}
(D_{\nu}F_{\gamma\rho}^{0})^{1} := D_{\nu}F_{\gamma\rho}^{1}+\frac{1}{2}\theta^{\alpha\beta}\{\partial_{\alpha}A_{\nu}^{0},\,\partial_{\alpha\beta}F_{\gamma\rho}^{0}\}.
\end{equation}
Using now Eq.~(\ref{eq:potential A second order}), the expression $\partial_{\gamma}A_{\rho}^{2}-\partial_{\rho}A_{\gamma}^{2}$ is calculated after a lengthy but straightforward procedure; we have
\begin{eqnarray}
\partial_{\gamma}A_{\rho}^{2}-\partial_{\rho}A_{\gamma}^{2} &=&  -\frac{1}{2}\theta^{\mu\nu}\left( A_{\mu}^{0}\partial_{\nu}F_{\gamma\rho}^{1}+A_{\mu}^{1}\partial_{\nu}F_{\gamma\rho}^{0}-F_{\gamma\mu}^{0}F_{\nu\rho}^{1} \right.
\nonumber \\[4pt]
&&-F_{\gamma\mu}^{1}F_{\rho\nu}^{0} + \frac{1}{2}\theta^{\alpha\beta}\partial_{\alpha}A_{\nu}^{0}\partial_{\alpha\beta}F_{\gamma\rho}^{0}
\nonumber \\[4pt]
&&\left. +2(\partial_{\mu}A_{\gamma}^{1}\partial_{\nu}A_{\rho}^{0}+\partial_{\mu}A_{\gamma}^{0}\partial_{\nu}A_{\rho}^{1})\right).
\end{eqnarray}
If we compare this equation with the field strength tensor given in Eq.~(\ref{eq:field strength-second order}), we get the following expression 
\begin{equation}
F_{\gamma\rho}^{2}=\partial_{\gamma}A_{\rho}^{2}-\partial_{\rho}A_{\gamma}^{2}+\theta^{\mu\nu}(\partial_{\mu}A_{\gamma}^{1}\partial_{\nu}A_{\rho}^{0}+\partial_{\mu}A_{\gamma}^{0}\partial_{\nu}A_{\rho}^{1}),
\end{equation}
which is very similar to Eq.~(\ref{fmunu1st}) obtained previously to first order.

\subsection{$n$-th order solutions}

Based on the previous results for the first two order solutions derived from the SW map, the following general structure of the solutions can be proposed~\cite{Ulker:2007fm} 
\begin{eqnarray}
\varLambda_{\lambda}^{n+1} & = & -\frac{1}{4(n+1)}\theta^{\mu\nu}\underset{p+q+r=n}{\sum}\{A_{\mu}^{p},\:\partial_{\nu}\varLambda_{\lambda}^{q}\}_{*r},
\label{eq:parameter n+1 order}
\\[4pt]
A_{\gamma}^{n+1} & = & -\frac{1}{4(n+1)}\theta^{\mu\nu}\underset{p+q+r=n}{\sum}\{A_{\mu}^{p},\:\partial_{\nu}A_{\gamma}^{q}+F_{\nu\gamma}^{q}\}_{*r}.
\end{eqnarray}
They are recursive relations for the noncommutative fields and by doing calculations similar to the previous ones, we can rewrite the $n$-th order term of the field strength as
\begin{eqnarray}
F_{\gamma\rho}^{n+1}&=&-\frac{1}{4(n+1)}\theta^{\mu\nu}\underset{p+q+r=n}{\sum}\left(\{A_{k}^{p},\,\partial_{l}F_{\gamma\rho}^{q}+(D_{l}F_{\gamma\rho})^{q}\} \right.
\nonumber \\[4pt]
&&\left. -2\{F_{\gamma k}^{p},\,F_{\rho l}^{q}\}_{*^r}\right),
\end{eqnarray}
where 
\begin{equation}
(D_{l}F_{\gamma\rho})^{n} := D_{l}F_{\gamma\rho}^{n}-i\underset{p+q+r=n}{\sum}[A_{l}^{p},\:F_{\gamma\rho}^{q}]{}_{*^r}.
\end{equation}
Here the sum is over all the values of $p$, $q$ and $r$ such that $p+q+r=n$; the subscript $*^r$ in a commutator $[f,g]_{*^r}$ means that we only consider the contributions of the form
\begin{eqnarray}
f(x)*^{r}g(x)&=&\dfrac{1}{r!}\left(\dfrac{i}{2}\right)^{r}\theta^{\mu_{1}\nu_{1}}\cdots\theta^{\mu_{r}\nu_{r}}
\nonumber \\[4pt]
&&\times \partial_{\mu_{1}}\cdots\partial_{\mu_{r}}f(x) \partial_{\nu_{1}}\cdots\partial_{\nu_{r}}g(x).
\end{eqnarray}
It is possible to find the same solution from a differential equation introduced in the original paper~\cite{Seiberg:1999vs}. This equation is often called the SW-differential equation and is obtained by varying the deformation parameter infinitesimally $\theta\rightarrow\theta+\delta\theta$; the solution of the SW differential equation to all orders has been obtained previously in~\cite{Bichl:2001cq}.

These solutions admit homogeneous contributions with arbitrary coefficients. This was noted previously by various authors in~\cite{Goto:2000zj,Jurco:2001rq,Moller:2004qq,Ettefaghi:2007cc} because the second order fields admitted different solutions; more recently this has been discussed in~\cite{Alboteanu:2007bp,Trampetic:2007hx}. Therefore, the most general solution should include these homogeneous terms; however, the main drawback in doing so is that recursive relations are more difficult to obtain to all orders.

\section{DQC in the Wu-Yang approach \label{sec:DQC-in-theWu-Yang approach} }
\label{secc:4}

In Dirac's original paper~\cite{Dirac:1931kp}, a singular solution of Maxwell's equations represents a magnetic monopole and the Dirac string is defined as the line where the gauge potential $A_{\mu}$ for the magnetic field becomes singular. The string is not observable since it can be rotated by a gauge transformation; this gauge transformation is also singular. 

Dirac's solution has been generalized as the 't Hooft\textendash Polyakov monopole, where the field is smooth gauge potentials. The question of having something similar for the Dirac string was analyzed in~\cite{PhysRevD.12.3845}, where Wu and Yang found a smooth construction of Dirac's monopole by separating $\mathbb{R}^{3}/\{0\}$ into two overlapping hemispheres defined as follows
\begin{eqnarray*}
R^{N} & : & 0\leq\theta<\pi/2+\delta,\quad r>0,\quad 0\leq\phi\leq2\pi,
\\
R^{S} & : & \pi/2-\delta<\theta\leq\pi,\quad r>0,\quad 0\leq\phi\leq2\pi.
\end{eqnarray*}
$R^N$ and $R^S$ are the north and south hemispheres and in both hemispheres $ t\in(-\infty,\;\infty)$; we have thus an overlapping region  $R^{N}\cap R^{S}$.  Each hemisphere is parametrized by an independent set of coordinates; associated with them there are two potentials $A_{\mu}^{N}(x)$ and $A_{\mu}^{S}(x)$ that are singularity-free everywhere in the domains of their definition. Explicitly we have
\begin{eqnarray}
A_{t}^{N}=A_{r}^{N}=A_{\theta}^{N}=0, & \quad & A_{\phi}^{N}=\dfrac{g}{r\sin\theta}(1-\cos\theta),
\nonumber \\
A_{t}^{S}=A_{r}^{S}=A_{\theta}^{S}=0, & \quad & A_{\phi}^{S}=-\dfrac{g}{r\sin\theta}(1+\cos\theta).
\label{eq:original potentials of Wu and Yan}
\end{eqnarray}
These potentials are required to satisfy the following conditions: 
\begin{enumerate}
\item They are related by a gauge transformation in the overlapping region; 
\item The magnetic field is obtained from their curls; 
\item In their respective domains, both potentials are free of singularities.
\end{enumerate}
The gauge potentials in Eq.~(\ref{eq:original potentials of Wu and Yan}) are related by the gauge transformation 
\begin{eqnarray}
A_{\mu}\rightarrow A_{\mu}^{\prime} & = & A_{\mu}+ie^{2ige\phi}\partial_{\mu}e^{-2ige\phi},\nonumber \\
 & = & A_{\mu}+\partial_{\mu}\lambda(x).\label{eq:gauge transformation}
\end{eqnarray}
in the overlapping region $R^{N}\cap R^{S}$ with corresponding gauge transformation
\begin{equation}
\lambda(x)=2g\phi=2g\arctan(y/x).
\end{equation}
It is a single-valued function only if 
\begin{equation}
2ige=\textrm{integer}=N.
\label{eq:quantization condition}
\end{equation}
Eq.~(\ref{eq:quantization condition}) is known as Dirac's quantization condition (DQC)~\cite{Dirac:1931kp}. 

\subsection{Wu-Yang procedure in Moyal space-time \label{sub:Wu-Yang-procedure-in}}

The basic ideas for the generalization of the above result to noncommutative spacetime will be briefly discussed here. First, we look for noncommutative potentials $A_{\mu}^{N}(x)$ and $A_{\mu}^{S}(x)$, such that the following happens:
\begin{enumerate}
\item The potentials in the overlapping region are related by the gauge transformation 
\begin{eqnarray}
A_{\mu}^{N/S}(x)\rightarrow A_{\mu}^{S/N}(x) &=& U* A_{\mu}^{N/S}(x)* U^{-1}
\nonumber \\[4pt]
&&+ iU*\partial_{\mu}U^{-1}.
\end{eqnarray}
where $U$ is an element of the noncommutative group $U_*(1)$.
\item Maxwell's equations with sources for the magnetic charges hold. 
\item There are no singularities in the potentials due to noncommutativity, the potentials remain free of singularities as in the classical case.
\end{enumerate}
The above conditions are similar to those mentioned in Sec.~\ref{sec:DQC-in-theWu-Yang approach} but adapted to the presence of noncommutativity. In particular, the second condition is imposed to find a relationship between the noncommutative gauge parameter and source terms. We will follow a perturbative treatment where the noncommutative gauge potential is written as
\begin{equation}
\hat A_{\mu}=A_{\mu}^{0}+A_{\mu}^{1}+A_{\mu}^{2}+\mathcal{O}(\theta^{3}),
\end{equation}
meanwhile the gauge parameter admits a similar writing
\begin{equation}
\hat \lambda=\lambda^{0}+\lambda^{1}+\lambda^{2}+\mathcal{O}(\theta^{3}).
\end{equation}
If the DQC is preserved then the noncommutative contributions to the gauge parameter should vanish.

\section{Noncommutative gauge transformations}
\label{secc:5}

The SW map is compatible with noncommutative gauge transformations of the form
\begin{equation}
\hat A_{\mu}\rightarrow \hat A_{\mu}^{\prime}=U* \hat A_{\mu}* U^{-1}+iU*\partial_{\mu}U^{-1},
\label{eq:gauge transformation-1}
\end{equation}
where $U \in U_{*}(1)$; the elements in this group are written as
\begin{equation}
U(x)=e_{*}^{i\lambda(x)}=1+i\lambda(x)+\frac{i^{2}}{2!}\lambda(x)*\lambda(x)+\ldots.
\end{equation}
The gauge group element up to second order in $\theta$ can be determined using previous results in the literature~\cite{salminen2011magnetic,LangvikSalminenTureanu2011}; its explicit form is
\begin{eqnarray}
e_{*}^{i\lambda(x)} & = & e^{-i\lambda(x)}+\frac{\theta^{pq}\theta^{kl}}{8}e^{-i\lambda(x)}\partial_{p}\partial_{k}\lambda
\nonumber \\[4pt]
&&\times \left(\frac{1}{2}\partial_{q}\partial_{l}\lambda+\frac{i}{3}\partial_{q}\lambda\partial_{l}\lambda\right)+\mathcal{O}(\theta^{3}).
\end{eqnarray}
On the other hand, the noncommutative gauge transformation Eq.~(\ref{eq:gauge transformation-1}) up to second order is given by
\begin{eqnarray}
A_{i}^{0} & \rightarrow & A_{i}^{0}+\partial_{i}\lambda(x),
\label{eq:zeroth transformation}
\nonumber \\[4pt]
A_{i}^{1} & \rightarrow & A_{i}^{1}-\theta^{kl}\partial_{k}\lambda(x)\partial_{l}A_{i}^{0}-\theta^{kl}\partial_{k}\lambda(x)\partial_{l}\partial_{i}\lambda(x),
\label{eq:first transformation}
\nonumber \\[4pt]
A_{i}^{2} & \rightarrow & A_{i}^{2}-\theta^{kl}\partial_{k}\lambda(x)\partial_{l}A_{i}^{1}+\dfrac{1}{2}\theta^{kl}\theta^{pq}\partial_{p}\lambda(x)
\nonumber \\[4pt]
&&\times \partial_{q}\left(\partial_{k}\lambda(x)\partial_{l}A_{i}^{0}+\frac{1}{3}\partial_{k}\lambda(x)\partial_{l}\partial_{i}\lambda(x)\right),
\label{eq:second transformation}
\end{eqnarray}
where we have used the same procedure given in \cite{salminen2011magnetic,LangvikSalminenTureanu2011}. To conclude, due to the first requirement in section \ref{sec:DQC-in-theWu-Yang approach}, we use Eqs.~(\ref{eq:second transformation}) to require that the following equations hold
\begin{eqnarray}
A_{i}^{N_{0}} & \rightarrow & A_{i}^{S_{0}}+\partial_{i}\lambda(x),
\nonumber \\[4pt]
A_{i}^{N_{1}} & \rightarrow & A_{i}^{S_{1}}-\theta^{kl}\partial_{k}\lambda(x)\partial_{l}A_{i}^{S_{0}}-\theta^{kl}\partial_{k}\lambda(x)\partial_{l}\partial_{i}\lambda(x)
\nonumber \\[4pt]
A_{i}^{N_{2}} & \rightarrow & A_{i}^{S_{2}}-\theta^{kl}\partial_{k}\lambda(x)\partial_{l}A_{i}^{S_{1}}+\dfrac{1}{2}\theta^{kl}\theta^{pq}  \partial_{p}\lambda(x)
\nonumber \\[4pt]
&&\times \partial_{q}\left(\partial_{k}\lambda(x)\partial_{l}A_{i}^{S_{0}}+\frac{1}{3}\partial_{k}\lambda(x)\partial_{l}\partial_{i}\lambda(x)\right).
\end{eqnarray}
In the following section we analyze the second requirement of Sec.~\ref{sec:DQC-in-theWu-Yang approach}, i.e. that the gauge potentials should satisfy
Maxwell's equations.

\section{Noncommutative Maxwell\textquoteright s equations in first and second order}
\label{secc:6}

A set of noncommutative Maxwell's equations for a static monopole 
\begin{eqnarray}
D_{\nu}* \widehat{F}^{\mu\nu} & = & 0,
\label{eq:"Amp=0000E8re's law"}
\\[4pt]
D_{\mu}*\mathcal{\widehat{F}}^{\mu\nu} & = & \widehat{J}_g^{\nu},
\label{eq:"Gauss's law"}
\\[4pt]
D_{\mu}*\widehat{J}^{\mu} & = & 0,
\label{eq:continuity equation}
\end{eqnarray}
was proposed in~\cite{LangvikSalminenTureanu2011}. Here 
\begin{eqnarray}
D_{\nu} & := & \partial_{\nu}-ie[\widehat{A}_{\mu},\,\cdot]_{*},
\nonumber \\[4pt]
\widehat{F}_{\mu\nu} & := & \partial_{\mu}\widehat{A}_{\nu}-\partial_{\nu}\widehat{A}_{\mu}-ie[\widehat{A}_{\mu},\,\widehat{A}_{\nu}]_{*},\end{eqnarray}
are the noncommutative covariant derivative and corresponding field strength tensor respectively. The dual field strength tensor is $\mathcal{\widehat{F}}_{\mu\nu} := \frac{1}{2}\epsilon^{\mu\nu\gamma\delta}\widehat{F}_{\gamma\delta}$. Eqs.~(\ref{eq:"Amp=0000E8re's law"}) and~(\ref{eq:"Gauss's law"}) are called Ampère's law and Gauss's law respectively; they are the analogs of the standard expressions in classical electrodynamics. Eq.~(\ref{eq:continuity equation}) is known as the continuity equation.

In constrast to~\cite{LangvikSalminenTureanu2011,salminen2011magnetic}, we do not consider the previous set of equations but a modification of it. Indeed, we first apply the SW map to the gauge potentials of the Wu-Yang approach to determine the corresponding noncommutative gauge potentials and using them, we will verify if the DQC is preserved order by order on the $\theta$ parameter; electric and magnetic sources, if any, will then be deduced from the modified Maxwell's equations.
 
Since we are considering a static monopole solution, we have $\widehat{J}_{g}^{i}=0$, i.e. there is no magnetic current. Also, $\widehat{J}_{g}^{0} \equiv\widehat{\rho}(r)=4\pi g\delta(r)+\rho^{1}(r)+\rho^{2}(r)+\mathcal{O}(\theta^{3})$ is the only nonvanishing component of the 4-dimensional noncommutative current $\widehat{J}^{\nu}$, giving rise to the total noncommutative magnetic charge
\begin{equation}
g_{NC} := \int J^{0}(x)d^{3}x
\end{equation}
This is a gauge invariant that can be calculated perturbatively; in the classical case it has the value $g$. In the static case we also note that the continuity equation is satisfied identically. In the following we move along the lines of~\cite{LangvikSalminenTureanu2011}.

\subsection{Ampère's law:} 

For a static solution, the electric field $E^i := F^{0i}=0$ vanishes, and all time-dependence is suppressed. In consequence, only the spatial components of Ampére's law provide non-trivial information. Let us focus then on $D_{k}*\widehat{F}^{ik}$, $i=1,2,3$. We have
\begin{eqnarray}
D_{k}*\widehat{F}^{ik} & = & \partial_{k}\widehat{F}^{ik}-i[\widehat{A}_{k},\,\widehat{F}^{ik}]_{*}
\nonumber \\[4pt]
 & = & \partial_{k}\left(F_{0}^{ik}+F_{1}^{ik}+F_{2}^{ik}\right) 
 \nonumber \\[4pt]
&& - i[A_{k}^{0}+A_{k}^{1}+A_{k}^{2},\,F_{0}^{ik}+F_{1}^{ik}+F_{2}^{ik}]_{*}
 \nonumber \\[4pt]
 & = & \partial_{k}F_{0}^{ik}+\partial_{k}F_{1}^{ik}+\partial_{k}F_{2}^{ik} +\theta^{pq}\partial_{p}A_{k}^{0}(\partial_{q}F_{0}^{ik}
 \nonumber \\[4pt]
&&+ \partial_{q}F_{1}^{ik}) +\theta^{pq}\partial_{p}A_{k}^{1}\partial_{q}F_{0}^{ik}+\mathcal{O}(\theta^{3}),
\end{eqnarray}
where $F_{jk}^{0},\,F_{jk}^{1}\,\textrm{and}\,F_{jk}^{2}$ are the field strength tensor components up to second order on $\theta$. Their explicit expressions are known from the SW map; using them we obtain
\begin{eqnarray}
D_{k}* F^{ik} &=& \partial_{k}(\partial^{i}A_{0}^{k}-\partial^{k}A_{0}^{i})+\partial_{k}(\partial^{i}A_{1}^{k}-\partial^{k}A_{1}^{i}
\nonumber \\[4pt]
 &&+\theta^{pq}\partial_{p}A_{0}^{i}\partial_{q}A_{0}^{k}) +\partial_{k}\left(\partial^{i}A_{2}^{k}-\partial^{k}A_{2}^{i} \right.
 \nonumber \\[4pt]
&&\left. +\theta^{pq}(\partial_{p}A_{1}^{i}\partial_{q}A_{0}^{k}+\partial_{p}A_{0}^{i}\partial_{q}A_{1}^{k})\right)
\nonumber \\[4pt]
&&+\theta^{pq}\partial_{p}A_{k}^{0}\partial_{q}\left(\partial^{i}A_{0}^{k}-\partial^{k}A_{0}^{i}+\partial^{i}A_{1}^{k}-\partial^{k}A_{1}^{i} \right.
\nonumber \\[4pt]
&&\left. +\theta^{rs}\partial_{r}A_{0}^{i}\partial_{s}A_{0}^{k}\right)+\theta^{pq}\partial_{p}A_{k}^{1}\partial_{q}\left(\partial^{i}A_{0}^{k}-\partial^{k}A_{0}^{i}\right)
\nonumber \\[4pt]
&&+\mathcal{O}(\theta^{3})
\nonumber \\[4pt]
 & = & \epsilon^{ikl}\partial_{k}B_{l}^{0}+\epsilon^{ikl}\partial_{k}B_{l}^{1}+\epsilon^{ikl}\partial_{k}B_{l}^{2}+\theta^{pq}
 \nonumber \\[4pt]
 &&\times \left(\partial_{k}(\partial_{p}A_{0}^{i}\partial_{q}A_{0}^{k})+\partial_{p}A_{k}^{0}\epsilon^{ikl}\partial_{q}B_{l}^{0}\right)
 \nonumber \\[4pt]
 && +\theta^{pq}\left(\partial_{p}A_{k}^{1}\epsilon^{ikl}\partial_{q}B_{l}^{0}+\partial_{p}A_{k}^{0}\epsilon^{ikl}\partial_{q}B_{l}^{1} \right.
 \nonumber \\[4pt]
&&+\partial_{k}(\partial_{p}A_{1}^{i}\partial_{q}A_{0}^{k}+\partial_{p}A_{0}^{i}\partial_{q}A_{1}^{k})+\theta^{rs}\partial_{p}A_{k}^{0}
\nonumber \\[4pt]
&&\left. \partial_{q}(\partial_{r}A_{0}^{i}\partial_{s}A_{0}^{k})\right)+\mathcal{O}(\theta^{3}),
\label{ampere2}
\end{eqnarray}
where we have defined $(\partial^{i}A_{n}^{k}-\partial^{k}A_{n}^{i}) =: \epsilon^{ikl}B_{l}^{n}$.

In the commutative case, Ampère law in its differential form is normally used to deduce the rotational of the magnetic field induced by an electric current but it can also be used in the opposite direction, namely to infer to electric current associated to a given magnetic field. According to this, let us write
 \begin{equation}
 D_{k}*\widehat{F}^{ik} = \dfrac{4\pi}{c}\widehat{J}_{e}^{i},
 \end{equation} 
 where $\widehat{J}_{e}^{i}=J_{e0}^{i}+J_{e1}^{i}+J_{e2}^{i}+\mathcal{O}(\theta^{3})$. Notice that we are thus allowing the presence of an electric current into Maxwell's equations. It follows from Eq,~(\ref{ampere2}) that
\begin{eqnarray}
(\nabla\times\vec{B}^{0})^{i} & = & J_{e0}^{i},
\nonumber \\
(\nabla\times\vec{B}^{1})^{i} & = & \theta^{pq}\left(\partial_{k}(\partial_{p}A_{0}^{i}\partial_{q}A_{0}^{k})+\partial_{p}A_{k}\epsilon^{ikl}\partial_{q}B_{l}^{0}\right) + J_{e1}^{i},
\nonumber \\
(\nabla\times\vec{B}^{2})^{i} & = & \theta^{pq}\left(\partial_{p}A_{k}^{1}\epsilon^{ikl}\partial_{q}B_{l}^{0}+\partial_{p}A_{k}^{0}\epsilon^{ikl}\partial_{q}B_{l}^{1} \right.
\nonumber \\[4pt]
&&+\partial_{k}(\partial_{p}A_{1}^{i}\partial_{q}A_{0}^{k}+\partial_{p}A_{0}^{i}\partial_{q}A_{1}^{k})
\nonumber \\[4pt]
&&\left. +\theta^{rs}\partial_{p}A_{k}^{0}\partial_{q}(\partial_{r}A_{0}^{i}\partial_{s}A_{0}^{k})\right) + J_{e2}^{i}, 
\label{eq:Amp=0000E8re's law zero, first and second order}
\end{eqnarray}
where $i,\,j\,,k=1,\,2,\,3$.

\subsection{Gauss's law:} 

We now proceed in the same way for Gauss' law. We have first that
\begin{eqnarray}
D_{i}*\mathcal{\hat F}^{i0} & = & \frac{1}{2}\epsilon^{i0jk}D_{i}*\hat{F}_{jk}=-\frac{1}{2}\epsilon^{ijk}D_{i}*\widehat{F}_{jk}
\nonumber \\
 & = & -\frac{1}{2}\epsilon^{ijk}\left(\partial_{i}\left(F_{jk}^{0}+F_{jk}^{1}+F_{jk}^{2}\right) \right.
 \nonumber \\[4pt]
 &&\left. -i[A_{i}^{0}+A_{i}^{1}+A_{i}^{2},\,F_{jk}^{0}+F_{jk}^{1}+F_{jk}^{2}]_{*}\right)
 \nonumber \\
 & = & -\frac{1}{2}\epsilon^{ijk}\left(\partial_{i}F_{jk}^{0}+\partial_{i}F_{jk}^{1}+\partial_{i}F_{jk}^{2} \right.
 \nonumber \\[4pt]
 &&\left. +\theta^{pq}\partial_{p}A_{i}^{0}(\partial_{q}F_{jk}^{0}+\partial_{q}F_{jk}^{1})+\theta^{pq}\partial_{p}A_{i}^{1}\partial_{q}F_{jk}^{0}\right)
 \nonumber \\[4pt]
 &&+\mathcal{O}(\theta^{3}).
 \label{eq:derivate of F}
\end{eqnarray}
Using Eqs.~(\ref{eq:field strength-first order}) and~(\ref{eq:field strength-second order}) into Eq.~(\ref{eq:derivate of F}), we can write Gauss' law as
\begin{eqnarray}
D_{i}*\mathcal{F}^{i0}-J^{0} & = & -\frac{1}{2}\epsilon^{ijk}\epsilon_{jkl}\partial_{i}B_{0}^{l}-4\pi g\delta(r)
\nonumber \\[4pt]
&&-\frac{1}{2}\epsilon^{ijk}\epsilon_{jkl}\partial_{i}B_{1}^{l}-\rho^{1}(x)
\nonumber \\[4pt]
&& -\frac{1}{2}\epsilon^{ijk}\epsilon_{jkl}\partial_{i}B_{2}^{l}-\rho^{2}(x)
 \nonumber \\[4pt]
 &&-\frac{1}{2}\epsilon^{ijk}\theta^{pq}\theta^{rs}\partial_{p}A_{i}^{0}\partial_{q}(\partial_{r}A_{j}^{0}\partial_{s}A_{k}^{0}),
 \label{eq:3erOrderGauss}
\end{eqnarray}
where we have used the fact that 
\begin{eqnarray}
&&\epsilon^{ijk}\theta^{pq}\partial_{p}A_{i}^{0}\partial_{q}F_{jk}^{0} = -\epsilon^{ijk}\theta^{pq}\partial_{i}(\partial_{p}A_{j}^{0}\partial_{q}A_{k}^{0}),
\nonumber \\[4pt]
&&\epsilon^{ijk}\theta^{pq}\left(\partial_{p}A_{i}^{0}\partial_{q}F_{jk}^{1}+\partial_{p}A_{i}^{1}\partial_{q}F_{jk}^{0}\right) = -2\epsilon^{ijk}\theta^{pq}\partial_{i}(\partial_{p}A_{j}^{1}\partial_{q}A_{k}^{0})
\nonumber \\[4pt]
&&+\epsilon^{ijk}\theta^{pq}\theta^{rs}\partial_{p}A_{i}^{0}\partial_{q}(\partial_{r}A_{j}^{0}\partial_{s}A_{k}^{0}),
\end{eqnarray}
together with the series expansion of $\hat J_g^0$. We further note that by a permutation of the indices, the last term in Eq.~(\ref{eq:3erOrderGauss}) vanishes. Since $\frac{1}{2}\epsilon^{ijk}\epsilon_{jkl}\partial_{i}B_{n}^{l}=\nabla\cdot\vec{B}^{n}$, we find the simple result
\begin{eqnarray}
&&\nabla\cdot\vec{B}^{0}=-4\pi g\delta(r), 
\nonumber \\[4pt]
&&\nabla\cdot\vec{B}^{1}=-\rho^{1}(r),
\nonumber \\[4pt]
&&\nabla\cdot\vec{B}^{2}=-\rho^{2}(r).
\label{eq:Gauss's law zero, first and second order}
\end{eqnarray}
These equations are similar to Eqs.~(\ref{eq:Amp=0000E8re's law zero, first and second order}) in that they allow us now to identify the sources of the modified monopole field.

\subsection{Combining Ampère's and Gauss's laws:}

Using the identity $\nabla^{2}\vec{B_{0}}=\nabla(\nabla\cdot\vec{B_{0}})+\nabla\times(\nabla\times\vec{B_{0}})$ we now combine Eq.~(\ref{eq:Gauss's law zero, first and second order}) and~(\ref{eq:Amp=0000E8re's law zero, first and second order}) in the usual way. We obtain then for the $i$-component
\begin{eqnarray}
\left(\nabla^{2}\vec{B_{0}}\right)^{i} & = & [\nabla(\nabla\cdot\vec{B_{0}})+\nabla\times(\nabla\times\vec{B_{0}})]^i
\nonumber \\[4pt]
 & = & \epsilon^{ijk}\partial_{j}\left(\epsilon^{klm}\partial_{l}B_{m}^{0}+\epsilon^{klm}\partial_{l}B_{m}^{1}+\epsilon^{klm}\partial_{l}B_{m}^{2} \right.
 \nonumber \\[4pt]
 &&+\theta^{pq}\left(\partial_{l}(\partial_{p}A_{0}^{k}\partial_{q}A_{0}^{l})+\partial_{p}A_{l}^{0}\epsilon^{klm}\partial_{q}B_{m}^{0}\right)
\nonumber \\[4pt]
 &  & +\theta^{pq}\left(\partial_{p}A_{l}^{1}\epsilon^{klm}\partial_{q}B_{m}^{0}+\partial_{p}A_{l}^{0}\epsilon^{klm}\partial_{q}B_{m}^{1} \right.
 \nonumber \\[4pt]
&& +\partial_{l}(\partial_{p}A_{1}^{k}\partial_{q}A_{0}^{l}+\partial_{p}A_{0}^{k}\partial_{q}A_{1}^{l})
\nonumber \\[4pt]
&& \left.\left. +\theta^{rs}\partial_{p}A_{l}^{0}\partial_{q}(\partial_{r}A_{0}^{k}\partial_{s}A_{0}^{l})\right)\right)
\nonumber \\[4pt]
 & = & (\delta^{il}\delta^{jm}-\delta^{im}\delta^{jl})\partial_{j}\partial_{l}\left(B_{m}^{0}+B_{m}^{1}+B_{m}^{2}\right)
 \nonumber \\[4pt]
 &&+\theta^{pq}\left(\epsilon^{ijk}\partial_{j}\partial_{l}(\partial_{p}A_{0}^{k}\partial_{q}A_{0}^{l}) \right.
 \nonumber \\[4pt]
 &&\left. +(\delta^{il}\delta^{jm}-\delta^{im}\delta^{jl})\partial_{j}(\partial_{p}A_{l}^{0}\partial_{q}B_{m}^{0})\right)
 \nonumber \\[4pt]
 &&+\theta^{pq}\left((\delta^{il}\delta^{jm}-\delta^{im}\delta^{jl})\partial_{j}(\partial_{p}A_{l}^{1}\partial_{q}B_{m}^{0} \right.
 \nonumber \\[4pt]
 && +\partial_{p}A_{l}^{0}\partial_{q}B_{m}^{1})+\epsilon^{ijk}\partial_{j}\partial_{l}(\partial_{p}A_{1}^{k}\partial_{q}A_{0}^{l}
 \nonumber \\[4pt]
 && +\partial_{p}A_{0}^{k}\partial_{q}A_{1}^{l})+\epsilon^{ijk}\partial_{j}\left(\theta^{rs}\partial_{p}A_{l}^{0} \right.
 \nonumber \\[4pt]
 &&\left. \left. \times \partial_{q}(\partial_{r}A_{0}^{k}\partial_{s}A_{0}^{l})\right)\right).
\end{eqnarray}
It is natural to ask whether Maxwell's equations and corresponding sources derived from the SW map are compatible with the perturbation
expansion approach used in~\cite{LangvikSalminenTureanu2011}. In this section we have calculated the equations of motion at zero, first and second order given by Eqs.~(\ref{eq:Amp=0000E8re's law zero, first and second order}) and~(\ref{eq:Gauss's law zero, first and second order}). It should be noted that the Maxwell equations derived here are compatible with the set of equations of motion given by Eqs.~(28), (31), (32) and Eqs. (23), (26), (27) of Refs.~\cite{LangvikSalminenTureanu2011,Langvik:2011tj} for the Ampère y Gauss laws respectively. However, contrary to~\cite{LangvikSalminenTureanu2011,Langvik:2011tj}, we have not assumed that the electric current vanishes and indeed we have the result that the SW map induces a non-vanishing electric current $\widehat{J_{e}}$.

\section{Solution of the noncommutative Maxwell\textquoteright s equations}
\label{secc:7}

We now use the SW map to determine the analytic expressions of the gauge potentials $A_{\mu}^{N}$and $A_{\mu}^{N}$ to all orders in the noncommutative parameter $\theta$. Both potentials satisfy Maxwell's equations with an appropriate source for the magnetic charge. In consequence, the criterion 2 in Sec.~\ref{sec:DQC-in-theWu-Yang approach} is satisfied for our potentials.

For comparison with previous results in the literature, we fix the values of $\theta^{\mu\nu}$ by imposing $\theta^{12}=-\theta^{21}$; all other components are set to zero. Furthermore, we use the original potentials of Wu and Yang, Eqs.~(\ref{eq:original potentials of Wu and Yan}), in cartesian coordinates. In the following $N^{0}$ and $S^{0}$ denote the zeroth order terms in $\theta$ in the northern and southern hemispheres respectively, and $r := \sqrt{x^{2}+y^{2}+z^{2}}$.

The gauge potentials to zero order in cartesian coordinates are
\begin{eqnarray}
A_{1}^{N0} & = & -y\dfrac{\left(r-z\right)}{\left(x^{2}+y^{2}\right)r}, \qquad A_{2}^{N0}=x\dfrac{\left(r-z\right)}{\left(x^{2}+y^{2}\right)r},
\nonumber \\[4pt]
A_{1}^{S0} & = & y\dfrac{\left(r-z\right)}{\left(x^{2}+y^{2}\right)r}, \qquad\quad A_{2}^{S0}=-x\dfrac{\left(r-z\right)}{\left(x^{2}+y^{2}\right)r},\nonumber \\[4pt]
A_{3}^{N0} & = & A_{3}^{N0}\,=\,A_{3}^{S0}\,=\,A_{0}^{N0}\,=\,A_{0}^{S0}\,=0.
\label{eq:Potential A0}
\end{eqnarray}
It is important to investigate the explicit form of the gauge potentials order by order to deduce a possible symmetry in the solution. For this purpose, we have calculated the spatial components of noncommutative gauge potential $\widehat{A}_{\mu}$ up to third order on perturbation theory explicitly and we have observed the following symmetry
\begin{eqnarray}
A_{k}^{(N/S){0}} &=& -\epsilon_{ki}x^{i}f_{0}^{(N/S)},
\nonumber \\[4pt]
A_{k}^{(N/S){1}} &=& -\epsilon_{ki}x^{i}f_{1}^{(N/S)},
\nonumber \\[4pt]
A_{k}^{(N/S){2}} &=& -\epsilon_{ki}x^{i}f_{2}^{(N/S)},
\nonumber \\[4pt]
A_{k}^{(N/S){3}} &=& -\epsilon_{ki}x^{i}f_{3}^{(N/S)},
\label{eq:S-WSolutions for A}
\end{eqnarray}
where $f_{0}^{(N/S)}, \dots,\:f_{3}^{(N/S)}$ are some functions having the following general structure $f_{k}^{(N/S)}=f_{k}^{(N/S)}(x^{2},\,y^{2},\,z)$. Based on this analysis of the first three order solutions, we can conjecture the general Ansatz 
\begin{equation}
A_{k}^{(N/S)_{n}} = -\epsilon_{ki}x^{i}f_{n}^{(N/S)},
\end{equation}
where $\epsilon_{ij}$ is the 2-dimensional Levi-Civita tensor, completely antisymmetric on its indices. The functions $f_{0}^{(N/S)}, \dots,\:f_{3}^{(N/S)}$ mentioned before have the following explicit expressions
\begin{widetext}
\begin{eqnarray}
f_{0}^{(N/S)} &=& \pm\dfrac{r\mp z}{r \rho^2},
\nonumber \\[4pt]
f_{1}^{(N/S)} &=& \pm\dfrac{\left(r\mp4z\right)\rho^4 + z^{2}\left(5r\mp6z\right)\rho^2 + 2z^{4}\left(r\mp z\right)}{2 r^{5} \rho^4},
\label{eq:f1}
\nonumber \\[4pt]
f_{2}^{(N/S)} &=& \pm\dfrac{\left(2r\mp13z\right)\rho^6 + z^{2}\left(17r\mp24z\right)\rho^4 + z^{4}\left(13r-15z\right)+\rho^4 \left(r\mp4z\right)+4z^{6}\left(r\mp z\right)}{2 r^8 \rho^6},
\nonumber \\[4pt]
f_{3}^{(N/S)} &=& \pm\dfrac{8z\left(13r\mp40z\right)\rho^8 + 2z^{3}\left(263r\mp370z\right)\rho^6 + 3y^{2}z^{5}\left(162r\mp203z\right)\rho^4 + 4z^{7}\left(65r\mp72z\right)\rho^2 + 56z^{9}\left(r\mp z\right)}{8 r^{10} \rho^8}.
\end{eqnarray}
\end{widetext}
where $\rho^2 := x^2 + y^2 = r^2 - z^2$ and the upper (lower) sign in the above expressions refers to the north (south) hemisphere.

An important criterion we should verify for the potentials is that they must be singularity-free. For this purpose it is convenient to find the components of the potential in spherical coordinates $(r, \vartheta, \phi)$; a straightforward calculation shows then that the potentials $A_{\mu}^{N}$ and $A_{\mu}^{S}$ are given by 
\begin{eqnarray}
A_{t}^{N_{m}}=A_{r}^{N_{m}}=A_{\vartheta}^{N_{m}}=0, & \quad & A_{\phi}^{N_{m}}=f_{m}^{N}\frac{y}{\sin\phi},
\nonumber \\
A_{t}^{S_{m}}=A_{r}^{S_{m}}=A_{\vartheta}^{S_{m}}=0, & \quad & A_{\phi}^{S_{m}}=f_{m}^{S}\frac{y}{\sin\phi},
\label{eq:potentials A-1}
\end{eqnarray}
where $m=0,\dots, 3$ means the perturbation order; the nonzero components can be written explicitly as
\begin{eqnarray}
A_{\phi}^{(N/S)_{0}} &=& \pm\frac{\tan\left(\frac{\vartheta}{2}\right)}{r},
\nonumber \\[4pt]
A_{\phi}^{(N/S)_{1}} &=& \pm\frac{[2\cos(\vartheta)+\cos(2\vartheta)]\tan\left(\frac{\vartheta}{2}\right)\sec^{2}\left(\frac{\vartheta}{2}\right)}{4r^{3}},
\nonumber \\[4pt]
A_{\phi}^{(N/S)_{2}} &=& \pm\frac{f(\vartheta) \tan\left(\frac{\vartheta}{2}\right)\sec^{4}\left(\frac{\vartheta}{2}\right)}{64r^{5}},
\nonumber \\[4pt]
A_{\phi}^{(N/S)_{3}} &=& \pm\frac{g(\vartheta) \tan\left(\frac{\vartheta}{2}\right)\sec^{6}\left(\frac{\vartheta}{2}\right)}{256r^{7}},
\end{eqnarray}
where
\begin{eqnarray}
f(\vartheta) &:=& 2\cos(\vartheta)+8\cos(2\vartheta)+14\cos(3\vartheta) + 5
\nonumber \\[4pt]
&&\times \cos(4\vartheta)+11,
\nonumber \\[4pt]
g(\vartheta) &:=& 46\cos(\vartheta)+32\cos(2\vartheta)+5\cos(3\vartheta) + 28
\nonumber \\[4pt]
&&\times \cos(4\vartheta) +29\cos(5\vartheta)+8\cos(6\vartheta) - 8.
\end{eqnarray}
From these expressions it is seen that in the limit where the polar angle vanishes, $\vartheta \rightarrow 0$, the components $A_{\phi}^{(N/S)_{m}} \rightarrow 0$. Therefore, both potentials $A_{\phi}^{N_{m}}$ and $A_{\phi}^{S_{m}}$ are singularity-free in their respective regions of validity; noncommutativity does not add divergences to the gauge potentials up to this order and we expect this to be a general feature.

We now use Eq.~(\ref{eq:S-WSolutions for A}) to find the sources associated to the Maxwell equations given in Sec.~\ref{secc:6}. If we insert Eq.~(\ref{eq:S-WSolutions for A}) into Eqs.~(\ref{eq:Amp=0000E8re's law zero, first and second order}) and~(\ref{eq:Gauss's law zero, first and second order}), we find that the sources are given by
\begin{eqnarray}
J_{e0}^{(N/S)i} & = & 0,
\nonumber \\[4pt]
J_{e1}^{(N/S)i} &=& \theta\,\epsilon^{ij3}x^{j}g_{1}^{(N/S)},
\nonumber \\[4pt]
J_{e2}^{(N/S)i} &=& \theta^{2}\,\epsilon^{ij3}x^{j}g_{2}^{(N/S)},
\label{ncelectriccurrent}
\end{eqnarray}
where
\begin{eqnarray}
g_{1}^{(N/S)} &:=& \mp\frac{3}{r^{6}},
\nonumber \\[4pt]
g_{2}^{(N/S)} &:=& \pm\frac{3\left[ \mp4z\left(2r\mp z\right)\rho^2 \pm z^{3}\left(r\mp z\right)+5\rho^4 \right]}{r^{10} \rho^2}.
\end{eqnarray}
We notice non-vanishing contributions to the electric current due to noncommutativity.

\subsection{The noncommutative parameter $\widehat{\varLambda}=\widehat{\varLambda}_{\lambda}\left(\lambda,\,A;\,\theta\right)$\label{sub:The-noncommutative-parameter}}

Having arrived to a general Ansatz for the noncommutative corrections to the gauge potentials of the magnetic monopole, we now proceed to discuss the DQC. The main point to be analyzed here is if the $\theta$-corrections to the standard gauge parameter $\lambda$ can be made to vanish when the gauge potentials are obtained from the SW map. From Eq.~(\ref{eq:parameter first order}) the noncommmutative parameter $\widehat{\varLambda}_{\lambda}$ to first order is
\begin{eqnarray}
\Lambda^{1}=-\dfrac{1}{2}\theta^{kl}A_{k}\partial_{l}\lambda.
\end{eqnarray}
The partial derivatives of the standard gauge parameter have a particular symmetry that is of great importance for the calculations that follow; they can be written as $\partial_{l}\lambda=-\epsilon_{lj}x^{j}g$, where $g :=\frac{2z}{x^{2}+y^{2}}$ and the corresponding potential is $A_{k}^{1} = -\epsilon_{ki}x^{i}f_1$ as pointed out before. Using these facts we compute the noncommutative correction $\Lambda^{1}$ to the gauge parameter obtaining
\begin{eqnarray}
\Lambda^{1} & = & -\dfrac{1}{2}\theta\epsilon^{kl}\left(-\epsilon_{ki}x^{i}f\right)\left(-\epsilon_{lj}x^{j}g\right) - \dfrac{1}{2}\theta\epsilon^{kl}\epsilon_{ki}x^{i}f\,
 \nonumber \\[4pt]
 &  &	\times \epsilon_{lj}x^{j}g -\dfrac{1}{2}\theta\delta_{i}^{l}x^{i}\,\epsilon_{lj}x^{j}fg - \dfrac{1}{2}\theta\,\epsilon_{ij}x^{i}x^{j}fg
 \nonumber \\[4pt]
&=&0,
 \label{eq:Lamda first order=00003D 0}
\end{eqnarray}
where we assume $\theta^{12}=-\theta^{21}=\theta$ as the only non-vanishing components. We see then that by using $\theta^{kl}=\theta^{12}\epsilon^{kl}=\theta\epsilon^{kl}$, the DQC to first order is preserved. The next step is to calculate the noncommutative second order correction $\Lambda^{2}$. From the explicit expression in Eq.~(\ref{eq:parameter second order}), we can write 
\begin{eqnarray}
\Lambda^{2} & = & -\dfrac{1}{4}\theta^{kl} \left( \left\{ A_{k}^{1},\,\partial_{l}\lambda\} + \{A_{k},\,\partial_{l}\Lambda^{1} \right\} \right).
\end{eqnarray}
The second term of this equation es zero, because we have already shown that $\Lambda^{1}=0$. Therefore, we only need to calculate $\{ \theta^{kl}A_{k}^{1},\,\partial_{l}\lambda \}$. From the previous section we have already computed the gauge potentials up to second order on $\theta$ and these can be written in general as $A_{k}^{2}=-\epsilon_{ki}x^{i}f_{2}$ where $f_{2}$ is given in Eq.~(\ref{eq:f1}). Taking this into account and following a procedure similar to the calculation of $\Lambda^{1}$, it is straightforward to show that the gauge parameter to second order also vanishes, i.e. $\Lambda^{2}=0$. 

\subsection{n\textendash th order $\Lambda$ :}

We now proceed to discuss the general case. First we recall the fact that the noncommutative corrections to the gauge parameter have the general form
\begin{eqnarray}
\Lambda^{n+1} & = & -\dfrac{1}{4(n+1)}\theta^{kl}\underset{_{p+q+r=n}}{\sum}\{A_{k}^{p},\,\partial_{l}\Lambda^{q}\}_{*r}.
\end{eqnarray} 
It is straightforward to see that the $\mathcal{O}(\theta^{r})$ contribution for the anticommutator $\{A_{k}^{p},\,\partial_{l}\Lambda^{q}\}_{*r}$, for $r$ an even number, vanishes; therefore, we can write
\begin{eqnarray}
\Lambda^{n+1} & = & -\dfrac{1}{4(n+1)}\theta^{\mu_{1}\nu_{1}}\left( \{ A_{\mu_{1}}^{n},\,\partial_{\nu_{1}}\lambda\}+\{A_{\mu_{1}}^{n-1},\,\partial_{\nu_{1}}\Lambda^{1}\} \right.
\nonumber \\[4pt]
&&\left. + ... +\{A_{\mu_{1}}^{n-s},\,\partial_{\nu_{1}}\Lambda^{s}\}+...+\{A_{\mu_{1}}^{0},\,\partial_{\nu_{1}}\Lambda^{n}\}\right)
\nonumber \\[4pt]
&  & -\dfrac{1}{4(n+1)}\theta^{\mu_{1}\nu_{1}}\{A_{\mu_{1}}^{0},\,\partial_{\nu_{1}}\Lambda\}_{*n}
\nonumber \\[4pt]
& = & -\dfrac{1}{4(n+1)}\theta^{\mu_{1}\nu_{1}}\left(\{A_{\mu_{1}}^{n},\,\partial_{\nu_{1}}\lambda\}+\{A_{\mu_{1}}^{n-1},\,\partial_{\nu_{1}}\Lambda^{1} \} \right.
\nonumber \\[4pt]
&&+...+\{A_{\mu_{1}},\,\partial_{\nu_{1}}\Lambda^{n}\} +\{A_{\mu_{1}}^{n-1},\,\partial_{\nu_{1}}\lambda\}_{*1}
\nonumber \\[4pt]
 &  & +...+\{A_{k}^{0},\,\partial_{\nu_{1}}\Lambda^{n-1}\}_{*1}
 \nonumber \\[4pt]
 &  & \qquad\hfill\:\vdots
 \nonumber \\[4pt]
 &  & \:+\{A_{\mu_{1}}^{1},\,\partial_{\nu_{1}}\lambda\}_{*n-1}+\{A_{k}^{0},\,\partial_{\nu_{1}}\Lambda^{1}\}_{*n-1}
 \nonumber \\[4pt]
 &  & \left.+\{A_{\mu_{1}}^{0},\,\partial_{\nu_{1}}\lambda\}_{*n}\right).
\end{eqnarray} 
Non-vanishing contributions of the anticonmutator $\{A,\,B\}_{*k}$ exists for $k$ even, where $A$ and $B$ are functions of $x^{i}$; therefore $\{A_{\mu_{1}}^{n-s},\,\partial_{\nu_{1}}\lambda\}_{*s}=0$ for $s$ odd and $\{A_{\mu_{1}}^{n-s},\,\partial_{\nu_{1}}\lambda\}_{*s}=2\,A_{\mu_{1}}^{n-s}*^{s}\partial_{\nu_{1}}\lambda$ for $s$ even. 

We have already seen that the noncommutative corrections $\Lambda^{1}$ and $\Lambda^{2}$ vanish. Let us now assume that this holds up to the $n$-th order, i.e. $\Lambda^{n}=0$; we would like to show that this assumption implies that the expression 
\begin{eqnarray}
\Lambda^{n+1} &=&  -\dfrac{1}{4(n+1)}\theta^{\mu_{1}\nu_{1}}\left(\{A_{\mu_{1}}^{n},\,\partial_{\nu_{1}}\lambda\} \right.
\nonumber \\[4pt]
&&\left. +\sum_{s=1}^n \{A_{\mu_{1}}^{n-s},\,\partial_{\nu_{1}}\lambda\}_{*s}\right)
\label{eq:Lamda (N+1); Lamnda N=00003D0}
\end{eqnarray}
for the $(n+1)$-th order also vanishes.

In the previous section we have conjectured the general Ansatz for the gauge potentials, namely $A_{\mu}^{n} =-\epsilon_{\mu i}x^{i}f_{n}$, where $f_n$ is some function. Assuming this and using a similar procedure as in Eq.~(\ref{eq:Lamda first order=00003D 0}) before, we obtain for the first term of Eq.~(\ref{eq:Lamda (N+1); Lamnda N=00003D0}) the result $\theta^{\mu_{1}\nu_{1}}A_{\mu_{1}}^{n}\partial_{\nu_{1}}\lambda=0$. For the second term we need to calculate the general expression $\theta^{\mu_{1}\nu_{1}}\{A_{\mu_{1}}^{n-s},\,\partial_{\nu_{1}}\lambda\}_{*s}$ for $s$ even only since, according to a previous remark, all the contributions with $s$ odd vanish. In the following we calculate this quantity for the most general case taking only into account the dependence on the coordinates of the gauge potentials and the gauge parameter.

We write then $\partial_{\nu_{1}}\lambda=-\epsilon_{\nu_{1}j}x^{j}g$ and $A_{\mu_{1}}=-\epsilon_{\mu_{1}i}x^{i}f$ together with $\theta^{\mu_{1}\nu_{1}}=\theta^{12}\epsilon^{\mu_{1}\nu_{1}}=\theta\epsilon^{\mu_{1}\nu_{1}}$, we have
\begin{eqnarray}
\theta^{\mu_{1}\nu_{1}}A_{\mu_{1}}*^{n}\partial_{\nu_{1}}\lambda &=& \dfrac{1}{n!}\left(\dfrac{i}{2}\right)^{n}\theta^{\mu_{1}\nu_{1}}\cdots\theta^{\mu_{n+1}\nu_{n+1}} 
\nonumber \\[4pt]
&&\times \partial_{\mu_{n+1}}\cdots\partial_{\mu_{2}}A_{\mu_{1}}\partial_{\nu_{n+1}}\cdots\partial_{\nu_{2}}\partial_{\nu_{1}}\lambda
\end{eqnarray}
where
\begin{eqnarray}
&&\partial_{\mu_{n+1}}\cdots\partial_{\mu_{2}}A_{\mu_{1}} =  -\epsilon_{\mu_{1}i}\left(\delta_{\mu_{2}}^{i}\partial_{\mu_{3}}\cdots\partial_{\mu_{n+1}}f+ \dots \right.
\nonumber \\[4pt]
&&\left. +\delta_{\mu_{n+1}}^{i}\partial_{\mu_{2}}\cdots\partial_{\mu_{n}}f+x^{i}\partial_{\mu_{2}}\cdots\partial_{\mu_{n+1}}f\right),
\nonumber \\[4pt]
&&\partial_{\nu_{n+1}}\cdots\partial_{\nu_{2}}\partial_{\nu_{1}}\lambda = -\epsilon_{\nu_{1}j}\left(\delta_{\nu_{2}}^{j}\partial_{\nu_{3}}\cdots\partial_{\nu_{n+1}}g+ \dots \right.
\nonumber \\[4pt]
&&\left. +\delta_{\nu_{n+1}}^{j}\partial_{\nu_{2}}\cdots\partial_{\nu_{n}}g+x^{j}\partial_{\nu_{2}}\cdots\partial_{\nu_{n+1}}g\right).
\end{eqnarray}
In consequence
\begin{widetext}
\begin{eqnarray}
\theta^{\mu_{1}\nu_{1}}A_{\mu_{1}}*^{n}\partial_{\nu_{1}}\lambda & = & \dfrac{1}{n!}\left(\dfrac{i}{2}\right)^{n} \theta \, \theta^{\mu_{2}\nu_{2}}\cdots\theta^{\mu_{n+1}\nu_{n+1}}[\left(\delta_{\mu_{2}}^{i}\partial_{\mu_{3}}\cdots\partial_{\mu_{n+1}}f+ \dots +\delta_{\mu_{n+1}}^{i}\partial_{\mu_{2}}\cdots\partial_{\mu_{n}}f\right)\epsilon_{ij}x^{j}\partial_{\nu_{2}}\cdots\partial_{\nu_{n+1}}g
\nonumber \\[4pt]
 &&+ \epsilon_{ij}x^{i}\partial_{\mu_{2}}\cdots\partial_{\mu_{n+1}}f\left(\delta_{\nu_{2}}^{j}\partial_{\nu_{3}}\cdots\partial_{\nu_{n+1}}g+ \dots +\delta_{\nu_{n+1}}^{j}\partial_{\nu_{2}}\cdots\partial_{\nu_{n}}g\right)+\epsilon_{ij}x^{i}x^{j}\partial_{\mu_{2}}\cdots\partial_{\mu_{n+1}}f\partial_{\nu_{2}}\cdots\partial_{\nu_{n+1}}g
 \nonumber \\[4pt]
 &&+ \left(\delta_{\mu_{2}}^{i}\partial_{\mu_{3}}\cdots\partial_{\mu_{n+1}}f+ \dots +\delta_{\mu_{n+1}}^{i}\partial_{\mu_{2}}\cdots\partial_{\mu_{n}}f\right)\epsilon_{ij}\left(\delta_{\nu_{2}}^{j}\partial_{\nu_{3}}\cdots\partial_{\nu_{n+1}}g+ \dots +\delta_{\nu_{n+1}}^{j}\partial_{\nu_{2}}\cdots\partial_{\nu_{n}}g\right).
\label{eq:matrix nxn}
\end{eqnarray}
The first two terms of this equation, that we denote as $S_n$, can be written in the following form
\begin{eqnarray}
S_n &=& \dfrac{1}{n!}\left(\dfrac{i}{2}\right)^{n} \theta\, \theta^{\mu_{2}\nu_{2}}\cdots\theta^{\mu_{n+1}\nu_{n+1}} ( \epsilon_{\mu_{2\,}j}x^{j}\partial_{\mu_{3}}\cdots\partial_{\mu_{n+1}}f\partial_{\nu_{2}}[\partial_{\nu_{3}}\cdots\partial_{\nu_{n+1}}g]+ \dots +\epsilon_{\mu_{n+1\,}j}x^{j}\partial_{\mu_{2}}\cdots\partial_{\mu_{n}}f[\partial_{\nu_{2}}\cdots\partial_{\nu_{n}}]\partial_{\nu_{n+1}}g
\nonumber \\[4pt]
 &&+ \epsilon_{i\,\nu_{2}}x^{j}\partial_{\mu_{2}}[\partial_{\mu_{3}}\cdots\partial_{\mu_{n+1}}f]\partial_{\nu_{3}}\cdots\partial_{\nu_{n+1}}g+ \dots +\epsilon_{i\,\nu_{n+1}}x^{i}[\partial_{\mu_{2}}\cdots\partial_{\mu_{n}}]\partial_{\mu_{n+1}}f\partial_{\nu_{2}}\cdots\partial_{\nu_{n}}g),
\end{eqnarray}
or equivalently, by changing the dummy indices, 
\begin{eqnarray}
S_n = \dfrac{1}{n!}\left(\dfrac{i}{2}\right)^{n} \theta\, \theta^{\mu_{2}\nu_{2}}\cdots\theta^{\mu_{n+1}\nu_{n+1}}(\epsilon_{\mu_{2\,}j}x^{j}\partial_{\nu_{2}}[\partial_{\mu_{3}}\cdots\partial_{\mu_{n}}f\partial_{\nu_{3}}\cdots\partial_{\nu_{n}}g]+...+\epsilon_{\mu_{n+1\,}j}x^{j}\partial_{\nu_{n+1}}[\partial_{\mu_{2}}\cdots\partial_{\mu_{n}}f\partial_{\nu_{2}}\cdots\partial_{\nu_{n}}g]).
\end{eqnarray}
\end{widetext}
From this we deduce that
\begin{eqnarray}
S_n &=& \dfrac 1n \left(\dfrac{i}{2}\right) \theta \sum_{s=2}^{n+1} \theta^{\mu_{s}\nu_{s}}\epsilon_{\mu_{s\,}j}x^{j}\partial_{\nu_{s}}\left(f*^{n-1}g\right)
\nonumber \\[4pt]
&=& \dfrac i2\theta^2 \,x^j \partial_j \left(f*^{n-1}g\right).
\label{eq: n+1column  n+1row}
\end{eqnarray}
Therefore we have $n$ terms of the type $x^{\mu_{s}}\partial_{\nu_{s}}\left(f*^{n-1}g\right)$. The third term of (\ref{eq:matrix nxn}) is clearly zero because $\epsilon_{ij}$ is an antisymetric tensor and hence $\epsilon_{ij}x^{i}x^{j}=0$. The last term in Eq.~(\ref{eq:matrix nxn}) can be seen as a $n\times n$
matrix where its elements can be written as
\begin{widetext}
\begin{eqnarray}
\dfrac{1}{n!}\left(\dfrac{i}{2}\right)^{n} \theta\, \theta^{\mu_{2}\nu_{2}}\cdots\theta^{\mu_{n+1}\nu_{n+1}}\epsilon_{ij} & ( & \delta_{\mu_{2}}^{i}\delta_{\nu_{2}}^{j}\partial_{\mu_{3}}\cdots\partial_{\mu_{n+1}}f\partial_{\nu_{3}}\cdots\partial_{\nu_{n+1}}g+...+\delta_{\mu_{2}}^{i}\delta_{\nu_{n+1}}^{j}\partial_{\mu_{3}}\cdots\partial_{\mu_{n+1}}f\partial_{\nu_{2}}\cdots\partial_{\nu_{n}}g
\nonumber \\[4pt]
 &  & \vdots\qquad\ddots
 \nonumber \\[4pt]
 & + & \delta_{\mu_{n+1}}^{i}\delta_{\nu_{2}}^{j}\partial_{\mu_{2}}\cdots\partial_{\mu_{n}}f\partial_{\nu_{3}}\cdots\partial_{\nu_{n+1}}g+...+\delta_{\mu_{n+1}}^{i}\delta_{\nu_{n+1}}^{j}\partial_{\mu_{2}}\cdots\partial_{\mu_{n}}f\partial_{\nu_{2}}\cdots\partial_{\nu_{n}}g).
\end{eqnarray}
\end{widetext}
The elements on the diagonal inside the parenthesis are given by
\begin{equation}
\dfrac 1n \left(\dfrac{i}{2}\right) \theta \underset{_{s=2}}{\overset{_{n+1}}{\sum}}\theta^{\mu_{s}\nu_{s}}\epsilon_{\mu_{s}\nu_{s}}f*^{n-1}g=i\theta^2\, (f*^{n-1}g),
\label{eq:diagonal}
\end{equation}
where we have $n$ terms of the type $\theta^{\mu_{s}\nu_{s}}\epsilon_{\mu_{s}\nu_{s}}f*^{n-1}g$ and we have used the fact that $\theta^{\mu_{s}\nu_{s}}\epsilon_{\mu_{s}\nu_{s}}=\theta\epsilon^{\mu_{s}\nu_{s}}\epsilon_{\mu_{s}\nu_{s}}=2\theta$.
The rest of the elements of this matrix are
\begin{equation}
\dfrac 1n \left(\dfrac{i}{2}\right) \theta \underset{_{r\neq s=2}}{\overset{_{n+1}}{\sum}}\theta (f*^{n-1}g) = \dfrac {i\theta^2}{2n} (n{}^{2}-n) (f*^{n-1}g),
\label{eq:rest of the elements}
\end{equation}
where we have used $\theta^{\mu_{r}\nu_{r}}\epsilon_{\mu_{r}\nu_{s}}=\theta\,\delta_{\nu_{s}}^{\nu_{r}}$; therefore we have $n^{2}-n$ terms of the form $\theta (f*^{n-1}g)$. Combining Eqs.~(\ref{eq: n+1column  n+1row}), (\ref{eq:diagonal}) and~(\ref{eq:rest of the elements}), we derive the following recursive formula for Eq.~(\ref{eq:matrix nxn})
\begin{eqnarray}
\theta^{\mu_{1}\nu_{1}}A_{\mu_{1}}*^{n}\partial_{\nu_{1}}\lambda &=& \dfrac{i\theta^2}{2n}\left[(n^{2}+n)\,f*^{n-1}g \right.
\nonumber \\[4pt]
&&\left.+ n\:x^{j}\partial_{j}\left(f*^{n-1}g\right)\right].
\label{eq:recursive formula}
\end{eqnarray}
Then, for a given $n$ even, we have reduced the calculation of $\theta^{\mu_{1}\nu_{1}}A_{\mu_{1}}*^{n}\partial_{\nu_{1}}\lambda$ to that of $(f*^{n-1}g)$. 

In Sec.~\ref{secc:7} we have seen that the functions $f$ and $g$ are quadratic functions of the cartesian coordinates $x$ and $y$; in consequence $\partial_{i}f=x^{i}F$ and $\partial_{i}g=x^{i}G$, $i=1, 2$ with $F$ and $G$ some quadratic functions on $x$ and $y$. With these elements at hand we have
\begin{eqnarray}
f*^{m}g &=& \dfrac{1}{m!}\left(\dfrac{i}{2}\right)^{m}\theta^{\mu_{m}\nu_{m}}\cdots\theta^{\mu_{1}\nu_{1}}
 \nonumber \\[4pt]
 &&\times \partial_{\mu_{m}}\cdots\partial_{\mu_{2}}x_{\mu_{1}}F\partial_{\nu_{m}}\cdots\partial_{\nu_{2}}x_{\nu_{1}}G,
\end{eqnarray}
Furthermore, we also have
\begin{eqnarray}
&&\partial_{\mu_{m}}\cdots\partial_{\mu_{2}}\partial_{\mu_{1}} f = \left(\delta_{\mu_{1}\mu_{2}}\partial_{\mu_{3}}\cdots\partial_{\mu_{m}}F+ \dots \right.
\nonumber \\[4pt]
&&\left. +\delta_{\mu_{1}\mu_{m}}\partial_{\mu_{2}}\cdots\partial_{\mu_{m-1}}F+x_{\mu_1}\partial_{\mu_{2}}\cdots\partial_{\mu_{m}}F\right),
\nonumber \\[4pt]
&&\partial_{\nu_{m}}\cdots\partial_{\nu_{2}}\partial_{\nu_{1}}g = \left(\delta_{\nu_{1}\nu_{2}}\partial_{\nu_{3}}\cdots\partial_{\nu_{m}}G + \dots \right.
\nonumber \\[4pt]
&&\left. +\delta_{\nu_{1}\nu_{m}}\partial_{\nu_{2}}\cdots\partial_{\nu_{m-1}}G+x_{\nu_1}\partial_{\nu_{2}}\cdots\partial_{\nu_{m}}G\right).
\end{eqnarray}
Therefore
\begin{eqnarray}
f*^{m}g & = & \dfrac{1}{m!}\left(\dfrac{i}{2}\right)^{m}\theta^{\mu_{m}\nu_{m}}\cdots\theta^{\mu_{1}\nu_{1}}[\left(\delta_{\mu_{1}\mu_{2}}\partial_{\mu_{3}}\cdots\partial_{\mu_{m}}F \right.
\nonumber \\[4pt]
&&\left. + \dots +\delta_{\mu_{1}\mu_{m}}\partial_{\mu_{2}}\cdots\partial_{\mu_{m-1}}F\right)x^{\nu_{1}}\partial_{\nu_{2}}\cdots\partial_{\nu_{m}}G
 \nonumber \\[4pt]
 &&+ \left(\delta_{\nu_{1}\nu_{2}}\partial_{\nu_{3}}\cdots\partial_{\nu_{m}}G+ \dots +\delta_{\nu_{1}\nu_{m}}\partial_{\nu_{2}}\cdots\partial_{\nu_{m-1}}G\right)
 \nonumber \\[4pt]
 &&\times x^{\mu_{1}}\partial_{\mu_{2}}\cdots\partial_{\mu_{m}}F+x^{\mu_{1}}x^{\nu_{1}}\partial_{\mu_{2}}\cdots\partial_{\mu_{m}}F
 \nonumber \\[4pt]
 &&\times \partial_{\nu_{2}}\cdots\partial_{\nu_{m}}G + \left(\delta_{\mu_{1}\mu_{2}}\partial_{\mu_{3}}\cdots\partial_{\mu_{m}}F + \dots \right.
 \nonumber \\[4pt]
 &&\left. +\delta_{\mu_{1}\mu_{m}}\partial_{\mu_{2}}\cdots\partial_{\mu_{m-1}}F\right) \left(\delta_{\nu_{1}\nu_{2}}\partial_{\nu_{3}}\cdots\partial_{\nu_{m-1}}G \right.
 \nonumber \\[4pt]
 &&\left.+ \dots +\delta_{\nu_{1}\nu_{m}}\partial_{\nu_{2}}\cdots\partial_{\nu_{n}}G\right).
\end{eqnarray}
Following a similar algebraic manipulation as that employed in Eq.~(\ref{eq:matrix nxn}), we can derive the following recursive formula for $f*^{m}g$
\begin{eqnarray}
f*^{m}g &=&\dfrac{\theta^2}{m(m-1)}\left(\dfrac{i}{2}\right)^{2}[ (m^{2}-m)\,F*^{m-2}G
\nonumber \\[4pt]
&&+(m-1)\:x^{j}\partial_{j}\left(F*^{m-2}G\right)].
\end{eqnarray}
It is important to note that this recursive relation is valid for any functions $f$, $g$, $F$ and $G$ that are quadratic functions in $x$ and $y$ and related by $\partial_{i}f=x^{i}F$ and $\partial_{i}g=x^{i}G$, $i=1, 2$. 

If in the previous expression we set $m=n-1$, we obtain
 \begin{eqnarray}
f*^{n-1}g &=&\dfrac{\theta^2}{(n-1)(n-2)}\left(\dfrac{i}{2}\right)^{2}[ (n^{2}-3n+2)
\nonumber \\[4pt]
&&\times F*^{n-3}G + (n-2)\:x^{j}\partial_{j}\left(F*^{n-3}G\right)].
\label{eq: recursive-formulaFG}
\end{eqnarray}

The result we have just obtained tell us that all we need to know is the value of $f*^{1}g$ in order to determinate $f*^{n-1}g$ for $n$ even. If we put first $n=2$ into Eq.~(\ref{eq: recursive-formulaFG}), we obtain the simple result $f*^{1}g=0$ and therefore we deduce that $f*^{n-1}g=0$ for $n$ even. Using this fact into Eq.~(\ref{eq:recursive formula}), we obtain then $\theta^{\mu_{1}\nu_{1}}A_{\mu_{1}}*^{n}\partial_{\nu_{1}}\lambda = 0$; this in turn implies that the second term in Eq.~(\ref{eq:Lamda (N+1); Lamnda N=00003D0}) also vanishes. In consequence $\Lambda^{n+1}=0$ for all $n$ and therefore $\widehat{\Lambda} = \lambda^0$. The DQC is then preserved under noncommutative corrections coming from gauge potentials obtained via the SW map.

\section{Conclusions}

In this work we have investigated the validity of DQC within the framework of noncommutative gauge theories. To do so, we have used the SW to define noncommutative gauge potentials associated to the commutative ones using as seeds the potentials introduced by Wu and Yang for the magnetic monopole.

With the noncommutative gauge potentials at hand, we have written down modified Maxwell's equations similar to those proposed previously in the literature. We differ however from previous treatments in that the gauge potentials are used to deduced the sources that should be present in these equations. The main difference we obtain from using this point of view is that an electric current also contributes to the curl of the noncommutative magnetic field; this is clearly seen when a perturbative series expansion in terms of the noncommutative parameter is considered and the corrections to the curl of the magnetic field are calculated. 

The SW map allows us first to give explicit expressions up to third order in perturbation theory; with this insight we have arrived to a general Ansatz for the gauge potentials to arbitrary order on the noncommutative parameter. The corresponding noncommutative gauge potentials are shown to be non-singular in their respective domains and this fact is explicitly verified up to third order on perturbation theory. Once these potentials are known, then the noncommutative contributions to the gauge parameter are calculated. For this, an iterative procedure derived directly from the iterative solution to the SW map has proven to be helpful. 

We have shown explicitly that corrections to the gauge parameter up to second order on the noncommutative parameter vanish. To prove this for all orders in perturbation theory, we have considered the general $n$-th order noncommutative contribution to the gauge parameter. Using the symmetries of the gauge potentials, we have been able to show that all noncommutative corrections vanish and that indeed DQC remains valid to all orders. 

This result is connected to the fact that an electric current arises due to noncommutativity. From Eqs.~(\ref{ncelectriccurrent}) we see that in the commutative limit, $\theta \to 0$, it vanishes and we recover the standard Maxwell's equations describing a magnetic monopole.

\acknowledgments
D. M.-C. acknowledges support from UAM Fellowship 2112800087.


\end{document}